\documentstyle[12pt]{article}

\begin{document}
\bibliographystyle{unsrt}

\begin{flushright} UMD-PP-94-95

\today
\end{flushright}

\vspace{6mm}

\begin{center}

{\Large \bf An  SO(10) $\times$ S$_{4}$ Scenario for Naturally Degenerate
Neutrinos\footnote{Work supported by the National Science Foundation Grant
PHY-9119745}}\\ [6mm]
\vspace{45mm}

{\bf Dae-Gyu Lee and R. N. Mohapatra}\\
{\it Department of Physics, University of Maryland\\  College Park,
Maryland 20742}\\ [4mm]

\vspace{20mm}

\end {center}

\begin{abstract}
The simplest scenario for the three known light neutrinos that fits the solar
and atmospheric
neutrino deficit and a mixed dark matter (MDM) picture of the universe requires
them to be
highly degenerate with $m_{\nu} \sim$ 1 - 2 eV.
We propose an SO(10) grand unified model with an S$_{4}$-horizontal symmetry
that
leads naturally to such a scenario.
 An explicit numerical analysis of the quark and lepton
sector of the model shows that it can lead to desired mass differences
to fit all data only for the small angle non-adiabatic MSW solution
to the solar neutrino puzzle.

\end{abstract}

\newpage
In the two recent papers, it has been pointed out by Caldwell and one of the
authors
(R.N.M.)\cite{CM:93} that if the existing data on solar\cite{SN} and
atmospheric
neutrino\cite{AN} deficit is confirmed and
 if the presently popular mixed dark matter (MDM)\cite{HDM}
model of the universe requiring a few eV neutrino as its hot component
is taken seriously,
 then the simplest three-light neutrino ($\nu_{e}$, $\nu_{\mu}$,
$\nu_{\tau}$) mass matrix that fits all data has the following form:

\begin{eqnarray}
M = \left( \begin{array}{ccc}
m+\delta_{1}s_{1}^{2} &-\delta_{1}c_{1}c_{2}s_{1} &-\delta_{1}c_{1}s_{1}s_{2}
\\
-\delta_{1}c_{1}c_{2}s_{1} & m+\delta_{1}c_{1}^{2}c_{2}^{2}+\delta_{2}s_{2}^{2}
&(\delta_{1}-\delta_{2})s_{2}c_{2}\\ -\delta_{1}c_{1}s_{1}s_{2}
&(\delta_{1}-\delta_{2})s_{2}c_{2} & m+\delta_{1}s_{2}^{2}+\delta_{2}c_{2}^{2}

 \end{array} \right), \end{eqnarray}
where $m$=2 eV, s$_{1}\sim.05$, s$_{2}\sim.38$,
 $\delta_{1}\sim 1.5 \times 10^{-6}$ eV
and $\delta_{1}\sim$ .2 to .002 eV.
 Here we have assumed that the solar neutrino puzzle is
solved by the small angle MSW (Mikheyev-Smirnov-Wolfenstein)\cite{MSW}
 solution\cite{HL}.
Note the high degree of degeneracy among the three neutrino species. An
immediate implication of this is that if the neutrinos are Majorana
particles , the neutrinoless double
beta decay would be measurable in the current generation experiments
involving $^{76}Ge$\cite{GE:93} and $^{130}Te$\cite{KK:94} thereby providing a
test of the
degenerate neutrino hypothesis.

It is then perhaps not premature to search for
 gauge models which can generate this highly degenerate
neutrino spectrum\cite{HS} in a technically natural manner. Such degeneracy
is suggestive of a horizontal symmetry, which will contain all
three neutrinos in one irreducible representation. This symmetry however
must be broken in the charged lepton sector.
 Since the see-saw mechanism\cite{seesaw}
generally connects the neutrino masses with the charged fermion masses
, the horizontal symmetry breaking in the
charged fermion sector  will cause  mass differences
between the neutrinos. Such mass differences are of course required and
from equation (1), we see that these must be
 very tiny. A simple scaling argument then says that the see-saw
scale must be in the range of $10^{12}$ GeV or so. Such mass scales
have their natural place in grand unified theories and
we will consider SO(10) as the flavor grand unifying theory in order
to understand the required neutrino spectrum.

Turning now to the horizontal symmetry, an obvious choice
is to consider  it to be  $SU(2)_H$
as has been done in several recent papers\cite{CM:93,HS}.
In this letter, we consider a somewhat more economical group
based on the permutation group $S_4$ as our horizontal symmetry.
We will assume that the symmetry is softly broken so that there is
no domain wall problem. We find that in this model, the fermion
sector is completely specified by fifteen arbitrary parameters,
twelve of which are fixed by the charged fermion sector ( i.e.
by six quark masses, three CKM angles and three charged lepton masses).
The structure of the neutrino mass matrix is then such that
only the small angle MSW solution to the solar neutrino problem,

Before presenting the details of the model, let us first
  review the basic strategy given in Ref.~\cite{CM:93}.
It is well-known that when the conventional see-saw mechanism for
neutrino-masses\cite{seesaw} is implemented
 in gauge models such as SO(10) or the
left-right symmetric models,
 it gets modified to the following form\cite{MS:81}
\begin{eqnarray}
\left( \begin{array}{cc}
 f v_{L} & m_{v^{D}} \\  m_{v^{D}}^{T} & f v_{R}

 \end{array} \right), \end{eqnarray}
where

\begin{eqnarray}
v_{L}=\lambda {v_{wk}^2 v_{R} \over M_{P}^2 };
\end{eqnarray}
$v_{R}$ is the scale of SU(2)$_{R}$-breaking
 and $M_{P}$ is breaking scale of parity.
Therefore, unless special care is taken
 to break parity symmetry at a scale higher than the
SU(2)$_{R}$ or U(1)$_{B-L}$, $v_{L}\sim \lambda v_{wk}^2 / v_{R}$ (since
$v_{R}\sim
M_{P}$ ).
The light neutrino masses are then given by:
\begin{eqnarray}
m_{\nu}\simeq fv_{L}-{m_{\nu^D} f^{-1} m_{\nu^D}^T \over v_{R} }.
\end{eqnarray}

Recall that the conventional see-saw formula omits the first term (which, as
just
mentioned, is justified only under special circumstances) leading to an
approximate
quadratic scaling relation between neutrino and up-quark mass (or in some
instances
charged lepton masses). We will however keep both the terms in the present
discussion.
Now notice that if due to some symmetry reasons,
 $f_{ab}=f_{0} \delta_{ab}$, then a degenerate
neutrino spectrum emerges.

Before discussing the fermion masses, let us briefly review some of the
discussions of
Ref.~\cite{CM:93}. We will consider the breaking of SO(10) $\rightarrow$
SU(2)$_L$
$\times$ SU(2)$_R$ $\times$ SU(4)$_C$ $\times$ P (denoted by $G_{224P}$) by
means of
a $\{{\bf 54}\}$-dim. Higgs multiplet. This symmetry is subsequently broken
down to the
standard model by a $\{{\bf 126}\}$-dim. Higgs multiplet. Detailed two-loop
analysis of the
mass scales in this model\cite{2step} leads to $v_{R}\sim 10^{13.6}$ GeV. So
that for $f_{0}
\lambda \sim 1/2$, we get
$f_{0}v_{L}\sim 1$ eV, as desired.

Let us now turn to the implication of S$_4$-symmetry for the charged fermion
and neutrino
masses, which is the main contribution of this paper. S$_4$-symmetry has been
used
before in the discussion of charged fermion masses at the electroweak
level\cite{PS:79}. It
has irreducible representations with dimension $\{{\bf 3}\}$, $\{{\bf
3^{'}}\}$, $\{{\bf 2}\}$,
$\{{\bf 1}\}$, and $\{{\bf 1^{'}}\}$. Our assignment of fermions and Higgs
multiplets to
irreducible representations of S$_4$ are shown in Table I.

The S$_4$-invariant Yukawa coupling can be written symbolically as
\begin{eqnarray}
L_{Y} &=& {1 \over \sqrt{3}}(\Psi_1 \Psi_1 + \Psi_2 \Psi_2 +\Psi_3 \Psi_3) (
h_{0} H_{0} + f \bar{\Delta}_0) \nonumber \\
& & +  {1 \over \sqrt{2}} \left[ (\Psi_3 \Psi_2 + \Psi_2 \Psi_3) (h_{2} H_{1} +
f_2
\bar{\Delta}_1)+ (\Psi_3 \Psi_3 + \Psi_2 \Psi_2 - 2 \Psi_1 \Psi_1) (h_{2} H_{2}
+ f_2
\bar{\Delta}_2) \right] \nonumber \\
& & + {f_3 \over \sqrt{2}} \left[ (\Psi_1 \Psi_3 + \Psi_3 \Psi_1) H_{3}  +
(\Psi_2 \Psi_1 +
\Psi_1 \Psi_2) H_4 +(\Psi_3 \Psi_3 - \Psi_2 \Psi_2) H_5  \right]  \nonumber \\
& & + H.~c.
\end{eqnarray}

We then assume that the S$_4$-symmetry is softly broken by the masses of $H_i$
($i=0,1,...,5$) and $\Delta_i$ ($i=0,1,2$), so that their vacuum expectation
values are
arbitrary. We also assume a softly broken U(1)$_{PQ}$ symmetry so that the
complex $\{
{\bf 10} \}$'s have only one coupling to the fermions.
The $H_i$'s therefore have vev's given by $\kappa_i^u$ and $\kappa_i^d$.
 Turning now to the
$\Delta$'s, we choose only the (mass)$^2$ of $\Delta_0$ negative and
 large so that the
(${\bf 1,3,\overline{10}}$) submultiplet of it
(the numbers denote representation under the group $G_{224P}$) acquire a vev
$v_R$ that breaks $G_{224P}$ down to the standard model.
 The remaining two $\Delta$'s
($\Delta_{1,2}$) have large (~$M_U$) positive masses so that
 their ({\bf 2,2,15 }) components
acquire induced vev's of order of the electroweak scale without any fine
tuning\cite{BM:93} due to the presence of S$_4 \times$ U(1)$_{PQ}$ invariant
terms such
as $\Delta_0 \overline{\Delta}_0 \Delta_i H_i$ in their potential (
where i=1,2).
A term of the form $\Delta_0 \overline{\Delta}_0 \Delta_0 H_0$ also induces
non-zero
vev's to the ({\bf 2,2,15}) submultiplets of $\Delta_{0}$. We denote these
$\Delta$-vev's by
$v_i^u$, $v_i^d$ ($i=0,1,2$). Note that the $\Delta_0$ coupling in Eq.~(5)
leads after
symmetry breaking to the degenerate neutrino masses.

The charged fermion and Dirac-neutrino mass matrices can now be written as
follows.
\begin{eqnarray}
M_{u,ab} &=& m_{u,ab}^{(10)}+f_0 v^u \delta_{ab} +m_{u,ab}^{(126)} \nonumber \\
M_{d,ab} &=& m_{d,ab}^{(10)}+f_0 v^d \delta_{ab} +m_{d,ab}^{(126)} \nonumber \\
M_{l,ab} &=& m_{d,ab}^{(10)}-3 f_0 v^d \delta_{ab}-3 m_{d,ab}^{(126)} \nonumber
\\
M_{\nu^D,ab} &=& m_{u,ab}^{(10)}-3 f_0 v^u \delta_{ab}-3
m_{u,ab}^{(126)}\nonumber\\
M_{\nu, ab} &=& f_0v_L\delta_{ab}- \left(M^2_{\nu^D}\right)_{ab}/(f_0v_R),
\end{eqnarray}
where
\begin{eqnarray}
m_{u,ab}^{(10)} &=& \left( \begin{array}{ccc}
a_0 -2 a_2 & a_4 & a_3 \\ a_4 & a_0+a_2-a_5 & a_1 \\ a_3 & a_1 & a_0+a_2+a_5
 \end{array} \right), \\
m_{u,ab}^{(126)}+f_0v^u\delta_{ab} &=& \left( \begin{array}{ccc}
d_0 -2 d_2 & 0 & 0 \\ 0 & d_0+d_2 & d_1 \\ 0 & d_1 & d_0+d_2
 \end{array} \right),
\end{eqnarray}
where $a_i$'s and $d_i$'s are products of type $h \kappa^u$ and $f v^u$
respectively. Similar
matrices can be written down for $m_{d,ab}^{(10)}$ and $m_{d,ab}^{(126)}$ by
replacing
$a_i$'s by $b_i$'s , $\kappa^u$ by $\kappa^d$ and $d_i$'s by $e_i$'s. This way
of writing makes
it clear that there are a total of 18 parameters. However, we can choose
a basis in which the up-quark mass matrix is diagonal, thereby getting
rid of three parameters and we are left with fifteen parameters.
 We will now fit these parameters so that the
six quark masses,
 three charged lepton masses and three CKM angles are reproduced.
In order to proceed , we first extrapolate these parameters from
the electro-weak scale to the scale $M_R$, where their values are:
\begin{center}
\begin{eqnarray}
\begin{array}{ccc}
m_u=.0013923,  & m_c=.36322   & m_t=75.900, \nonumber \\
m_d=.0024297,  & m_s=.047775   & m_b=1.38975, \nonumber \\
s_{12}=\pm.2200,  & s_{13}=.00624   & s_{23}=.05200, \nonumber \\
m_e=.0004896,  & m_{\mu}=.10080   & m_{\tau}=1.71264.
\end{array}
\end{eqnarray}
\end{center}

We have chosen all masses to be positive.
We do the fitting as follows. In the basis where
  $M_u$ is diagonal, we have  $M_d=U_{ CKM} M_d^{(diag)}
U_{CKM}^{\dag}$.
The first choice leads to six constraints on the original ( before choosing
the up-quark basis )
 nine parameters leaving three arbitrary. We
choose them to be $a_0$, $a_1$, and $a_2$.
 Similarly the $M_d$ equation leaves three
arbitrary parameters chosen to be $b_0$, $b_1$, and $b_2$.
The $b_0$, $b_1$, and $b_2$ are then determined by three
equations that relate them to $Tr M_l$, $Tr M^2_l$ and $Tr M^3_l$
which of course are easily expressed in terms of
 the $m_e$, $m_{\mu}$, and
$m_{\tau}$ . The matrix $M_l$ is then completely determined.
  Diagonalizing it as usual i.e.
$M_l=U_{E} M_l^{(diag)} U_{E}^{\dag}$, where $M_l^{(diag)}=$ Diag($m_e$,
$m_{\mu}$,
$m_{\tau}$), the three angles that parameterize $V_{E}$ are  predicted.
In particular, note that, in the basis in which $M_u$ is diagonal, $\nu_e
\nu_{\mu}$ mixing
arises predominantly from the ($e, \mu, \tau$) sector due to the fact that
$M_{\nu^D}$ is block diagonal
 and fitting
$m_e$, $m_{\mu}$, and $m_{\tau}$ leads a prediction for
$\theta_{\nu_{e} \nu_{\mu}}$. We find that for negative $s_{12}$, two sets of
solutions for the mixing matrix $V_E$ correspond to the correct charged
lepton masses. Only one of them leads to the desired $\nu_e$-$\nu_{\mu}$
mixing angle and we choose this one ;
the parameters
 $a_0$, $a_1$, and $a_2$ not fixed by the up-quark sector, are then varied
 to
obtain the required mass difference squares between
 $m_{\nu_e}$, $m_{\nu_\mu}$, and $m_{\nu_\tau}$
 to fit the data.

Let us now briefly discuss the dependence of our solutions on the
three  parameters $a_0$,$a_1$ and $a_2$. Requiring $\Delta m^2_{\nu_\mu
-\nu_{\tau}}\simeq .1 eV^2$ implies that, $a_0+a_2\simeq 38 GeV$;
if we further assume that the non-degenarate part of the contribution
to the neutrino masses have a hierarchical structure between the
first and the second generation leading to $\Delta m^2_{\nu_e-{\nu_\mu}}
\simeq 10^{-6} eV^2$, this leads to $a_0\simeq 2a_2 \simeq 25.36 GeV$
and $a_1\leq 2$.
 In figure 1, we have shown the
dependence of the $\nu_e$- $\nu_{\mu}$ mass difference on $a_1$ for the
above choices of $a_2$ and $a_0$ and we see that
the desired range comes only for $a_1\leq 2$.
We give below the detailed solution for the neutrino masses and mixings
for this case.
(We have written $m_{\nu_i}=m_0+
m_{\nu_i}^{'}$, where $m_0$ is the direct $v_L$ contribution.)
\begin{eqnarray}
(m_{\nu_{e}}^{'}, m_{\nu_{\mu}}^{'}, m_{\nu_{\tau}}^{'})={1 \over  f v_R}
(-0.0000174465,
-0.129248, -5759.27 ) \mbox{GeV}^2,\nonumber
\end{eqnarray}
and

\begin{eqnarray}
V^l &=& \left( \begin{array}{ccc}
-.9982 & .05733 & .01476 \\ .05884 & .9334 & .3541 \\ -.006523 & -.3544 & .9351
\end{array} \right).
\end{eqnarray}

Note that, for $v_R \simeq 10^{13.6}$ GeV and $f \sim 3$, this predicts
$|m_{\nu_{\mu}}^{2}-m_{\nu_e}^{2}| \sim 4 \times 10^{-6}$ eV$^2$ for $m_0=$ 2
eV,
$|m_{\nu_{\tau}}^{2}-m_{\nu_{\mu}}^{2}| \sim .2$ eV$^2$, which are
 in the range required to
solve both the solar and atmospheric neutrino deficit for the values of
 $\theta_{\nu_{e}
\nu_{\mu}}$ and $\theta_{\nu_{\mu} \nu_{\tau}}$ given above.
In particular, we wish to note the preference of theory for the
small angle MSW solution to the solar neutrino problem. This
is quite interesting in view of the recent conclusions\cite{KIR}
that the large angle MSW solutions have a discouragingly large
$\chi^2$-fit.

 We wish to remark that our solutions do not depend on the choice
of sign for the up quark masses. We have checked that we
lose the desired values of masses
and mixings if we choose $m_s$ to be negative for both signs of the
Cabibbo angle.

In conclusion, we have constructed an SO(10) $\times$ S$_4$ model, which leads
to a
degenerate neutrino mass scenario including correct mixing angles required to
fit
atmospheric and solar neutrino data as well as the hot dark matter in the
universe. A
critical test of the model is the observation of a non-zero signal in
neutrinoless double data
decay in the current generation of $^{76}$Ge and $^{130}$Te
experiments. Also our model prefers only the small angle MSW solution to
the solar neutrino puzzle, a result which is already indicated in recent
analyses and will surely be tested once more data accumulates.

\newpage
\section*{Table Caption}
Table I: Transformation property of fermions and Higgs multiplets under
S$_4$-symmetry.
The $\{{\bf 10} \}$-dimensional multiplets are chosen to be complex..

\vspace*{10mm}
{\small
\begin{center}

\begin{tabular}{ |c|c| } \hline\hline
Multiplet & S$_4$ representation \\ \hline
Fermions  &             \\

$\Psi_{a}, a=1,2,3$         & $\{ {\bf 3} \}$  \\ \hline
Higgs Bosons &          \\
$\{ {\bf 126} \}$ \hspace{5.5mm} $\Delta_{0}$ \hspace{1.5mm} &  $\{ {\bf 1} \}$
 \\
$\{ {\bf 126} \}$ \hspace{5mm} $\Delta_{1,2}$  &  $\{ {\bf 2} \}$  \\
$\{ {\bf 10} \}$ \hspace{7mm} H$_{0}$  \hspace{1.5mm} &  $\{ {\bf 1} \}$  \\
$\{ {\bf 10} \}$ \hspace{6mm} H$_{1,2}$  \hspace{1.mm}&  $\{ {\bf 2} \}$  \\
$\{ {\bf 10} \}$ \hspace{5mm} H$_{3,4,5}$      &  $\{ {\bf 3} \}$ \\ \hline
\hline
\end{tabular}
\end{center}
\begin{center}
Table I.

\end{center}
}

\section*{Figure Captions}

\vspace{10mm}
\noindent Fig. 1:
$m^{\prime}_{\nu_e}$ and $m^{\prime}_{\nu_\mu}$
 are plotted as a function of  $a_1$ for  $a_0=2a_2=25.38 GeV$.
   The scale of the
vertical axes is 8.37 $\times$ 10$^{-6}$ eV for $v_R$ = 10$^{13.6}$ GeV
 and $f = 3$.

\end{document}